\documentclass{hotnets22}

\usepackage{times}  
\usepackage[hidelinks]{hyperref}
\usepackage{titlesec}

\hypersetup{pdfstartview=FitH,pdfpagelayout=SinglePage}

\usepackage{amsmath,amssymb,amsfonts}
\usepackage{algorithm}
\usepackage{algorithmic}
\usepackage{array} 
\newcolumntype{P}[1]{>{\centering\arraybackslash}m{#1}} 
\usepackage{booktabs}
\usepackage{graphicx}
\usepackage{textcomp}
\usepackage{xcolor}
\newcommand{\comment}[1]{\\ \texttt{\small{/* #1 */}}} 
\newcommand{\remove}[1]{} 

\usepackage[textfont=bf, labelfont=bf]{caption}
\begin{document}


\title{EdgeP4: A P4-Programmable Edge Intelligent Ethernet Switch for Tactile Cyber-Physical Systems}

\author{Nithish Krishnabharathi Gnani, Joydeep Pal, Deepak Choudhary, Himanshu Verma,\\
Soumya Kanta Rana, Kaushal Mhapsekar\textsuperscript{1}, T. V. Prabhakar, and Chandramani Singh\\
\textit{Indian Institute of Science, Bengaluru, India, ~~~~\textsuperscript{1} NC State University, Raleigh, USA}\\
\textit{\{nithishgnani, joydeeppal, deepakcl, himanshuv, soumyarana\}@iisc.ac.in, \textsuperscript{1} kmhapse@ncsu.edu}
}

\maketitle

\begin{abstract}

Tactile Internet based operations, e.g., telesurgery, rely on end-to-end closed loop control for accuracy and corrections. The feedback and control are subject to network latency and loss. We design two edge intelligence algorithms hosted at P4 programmable end switches. These algorithms locally compute and command corrective signals, thereby dispense the feedback signals from traversing the network to the other ends and save on control loop latency and network load. We implement these algorithms entirely on data plane on Netronome Agilio SmartNICs using P4. Our first algorithm, \textit{pose correction}, is placed at the edge switch connected to an industrial robot gripping a tool. The round trip between  transmitting force sensor array readings to the edge switch and receiving correct tip coordinates at the robot is shown to be less than $100~\mu s$. The second algorithm, \textit{tremor suppression}, is placed at the edge switch connected to the human operator. It suppresses physiological tremors of amplitudes smaller than $100~\mu m$ which not only improves the application's performance but also reduces the network load up to $99.9\%$. Our solution allows edge intelligence modules to seamlessly switch between the algorithms based on the tasks being executed at the end hosts.

\end{abstract}

\section{INTRODUCTION AND MOTIVATION} \label{introduction}

\begin{figure*}[t]
    \centering
    \includegraphics[width=\textwidth]{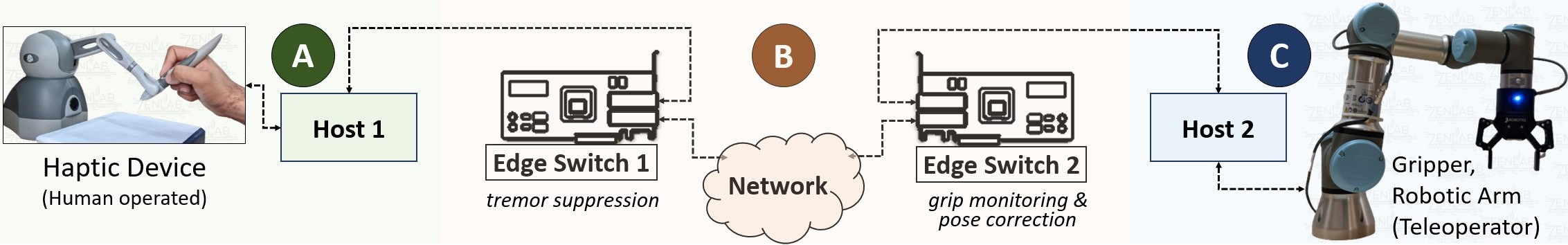}
    \caption{Cyber-Physical System Testbed}
    \label{testbed}
\end{figure*}

Grip inspection is an essential requirement in applications involving robotic arms with fingered grippers. In a traditional robot-assisted surgery and diagnostics, the human being has manual control over the robot. For example, in the DaVinci System \cite{freschi2013technical}, the doctor has to be colocated with the multi-arm robotic platform for manipulating it. This constraint can be overcome through Time-Sensitive Networking (TSN), which not only offers bounded latency and reliability over longer distances but also effectively addresses cyber-sickness in humans \cite{polachan2022assessing}. For example, consider the case of gripping a surgical tool where the teleoperator is expected to firmly grip the tool and perform a task. The human positions the robot around the tool, closes the gripper and inspects the grip. The pose of the ``tool centre point'' of the robot is provided as feedback to the human operator. If the grip is unsatisfactory, the human operator can open the gripper, re-position the robot, and close the gripper again. This process is iterated until a satisfactory grip is achieved. Several action-feedback cycles may be required before a firm hold on the tool by the robot is attained.

Edge intelligence systems can instead complete the gripping activity with human intervention and can provide feedback to her in real time. This would allow the operator to focus on the core application-related actions. Edge computing devices like the Nvidia Jetson Nano \cite{jetson_ref}, and Google Coral \cite{coral_ref}, designed for artificial intelligence (AI) and machine learning (ML), can handle such tasks in real time and meet the latency requirements of tactile cyber-physical systems. Interestingly, grip inspection is often performed using cameras, relying on multiple cameras to provide complete visibility when the gripper engulfs the tool. Synchronization of camera timestamps is required for capturing simultaneous snapshots. Furthermore, image processing algorithms are used to evaluate the grip and calculate corrective pose data.

In this work, we implement intelligence at the switch ports in a tactile cyber-physical system consisting of TSN-capable switches built using P4 (Programming Protocol independent Packet Processors) programmable Smart Network Interface Cards (SmartNIC), a haptic device, and a robotic arm equip-ped with a two-fingered gripper. This ensures that the robotic gripper, on receiving a command from the operator, inspects the grip and autonomously performs pose correction to grip the tool while simultaneously providing haptic feedback to the operator. This not only reduces latency but also allows more flexible management of different robots performing various tasks. In Fig. \ref{testbed}, the edge intelligence algorithm runs on \textit{Edge Switch 2}. While integrating intelligence into switch ports eliminates the need for additional hardware and synchronization, running AI-based inference on these ports, using image data from multiple cameras, can be challenging due to computationally expensive image processing. Moreover, communication of image data is required between the cameras and the switch. In contrast, our proposed solution utilizes force sensor arrays for grip inspection, which only requires the gripper to press against the tool for detection. We have developed a conventional geometry-based algorithm built to work with the sensor array data. Furthermore, we ensure that the human operator's hand tremors which are not desired in the robot's output motion are masked at the source (\textit{Edge Switch 1}). Thus, the switch port intelligently handles egress port traffic to improve the robot's performance and to reduce the network load at the same time. Our solution strikes a good balance between the use of edge computing modules and microcontroller-based modules that may not meet the stringent compute and latency demands. 


Our contributions are as follows: 
\begin{itemize}
    \item Designing and implementing an algorithm on the edge switch port that monitors the grip and provides pose correction to firmly hold a tool in position. \vspace{-5pt}
    \item Designing and implementing an intelligent hand tremor detection algorithm on the edge switch port to reduce its effect on the task performed by the robot as well as to reduce the network traffic. \vspace{-5pt}
    \item Designing a framework for switch ports to detect and automatically switch between several algorithms specific to a task. 
\end{itemize}

\section{RELATED WORK} \label{rel-work}

P4 is an open-source programming language used in data plane programming which includes  custom functionalities such as the forwarding process \cite{bosshart2014p4}. Prior work on usage of P4's match action functionality and its ability to invoke external MicroC functions is restricted to data packet inspection and inferencing AI models that are usually built offline. Pre-trained ML models can be mapped to programmable swi-tches for in-network classification at line rate via match action pipelines \cite{xiong2019switches}. Sankaran et al.~\cite{sankaran2021p4} implement decision trees on NetFPGA using if-else statements inside actions in P4 and store the classification in a packet header field. Langlet et al.~\cite{langlet2019towards} implement Artificial Neural Network inference into the data plane of P4 programmable SmartNICs with flow metadata extraction. Simpson and Pezaros~\cite{simpson2022revisiting} implement online learning in SmartNICs using reinforcement learning. Since large ML models can introduce significant latencies, in our work, the switch port runs our pose correction algorithm under in-situ real-time conditions. We thus introduce suitability for tactile applications.

Prior work on gripping tasks propose solutions based on vision, tactile information, and fuzzy logic \cite{van2019towards, sadeghian2022vision,muthusamy2020neuromorphic,veiga2018grip}. While  tactile sensors can detect slip, pose correction is accomplished using multiple camera systems. Not only this method introduces uncertainty in a stable grip, positioning multiple cameras for a comprehensive set of scenarios is also restrictive. Also, while most grippers are equipped to control the force given by their motors to the fingers, defining the grasping force \cite{gripper_force_ref}, there are no force sensor array solutions for estimating the correct gripping of the tools.


Hand tremors are rhythmic shaking movements of the hand that occur involuntarily. Alty et al.~\cite{alty2017use} mention different types of tremors, out of which physiological tremor is of our interest. This tremor is rarely visible to eyes and typically involves a fine shaking of hands and fingers. Prior studies indicate that physiological tremors typically start in a frequency range between $8-9~Hz$, with an amplitude (hand excursions) of about $100~\mu m$. Gradually over time, the amplitude can 
become as high as $2~mm$ in the frequency range of about $4-6~Hz$ \cite{stiles1976frequency}. There have been significant developments in detecting and modelling physiological tremors \cite{veluvolu2010double,zhang2005dsp}. A few works have applied these models in building microsurgical instruments \cite{ang2000active}\cite{riviere2003toward}. Since these instruments are used directly on patients, aggressive tremor filtering algorithms use piezoelectric actuators to counteract the tremor. Riviere et al.~\cite{riviere2003toward, riviere1998adaptive} propose weighted-frequency Fourier linear combiner (WFLC) for 1 DoF and 3 DoF, respectively. They demonstrate  $69\%$ reduction in tremors. On the other hand, robotic-assisted remote surgeries require algorithms to filter tremors as well as to block tremor data being communicated over the network. Our solution does not need additional actuators to arrest the tremors.

\section{SYSTEM ARCHITECTURE} \label{architecture}

\subsection{Testbed Description} \label{testbed-desc}
\begin{figure}[ht]
\vspace{-5pt}
\centerline{
\includegraphics[width=0.30\textwidth]{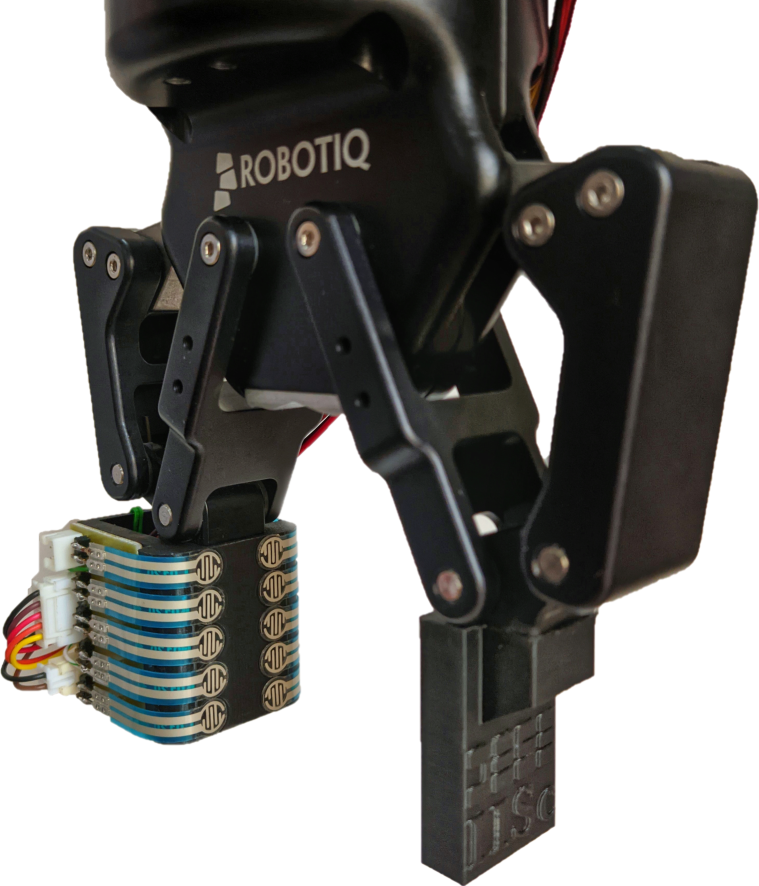}
}
\caption{Custom fingertips with force sensor array attached on the gripper}
\label{fsr-photo}
\end{figure}

Fig.~\ref{testbed} shows the testbed that we have constructed in-house. It consists of three domains - A.~{\it human operator domain}, B.~{\it  network domain} and C.~{\it teleoperator domain}. Domain A consists of a haptic-feedback device, Geomagic Touch \cite{GeoT_details}, connected to \textit{Host 1} and manipulated by a human operator. C-programming tools are employed to use the OpenHaptics toolkit running on \textit{Host 1} to obtain the kinematic data from Geometric Touch. This data is communicated to Domain C which consists of a UR3 (CB version) robotic arm from Universal Robots \cite{UR3_ref} equipped with a Robotiq 2F-85 gripper \cite{Robotiq_ref} connected to \textit{Host 2}. A maximum force of $235~N$ may be applied to grasp an object. Fig. \ref{fsr-photo} shows a custom-built $5 \times3$ array of force sensors with custom fingertips attached to the gripper. UR3 is designed for factory settings with a fixed set of tasks that can be programmed using their patented PolyScope interface. To achieve real-time control, we leverage the robot's ability to be externally controlled by a connected computer (\textit{Host 2}) running Robot Operating System (ROS) with Python programming in conjunction with PolyScope. We use the \textit{CartesianTrajectoryController} available in the \textit{rospy} library of the ROS driver of UR3 to command the robot to move to the desired 3D coordinates. The pose of the robot as well as the force sensor data are sent to Domain A as feedback. The hosts run Linux Ubuntu 20.04 operating system. In the teleoperation process, a human operator holds and manipulates the stylus of the haptic device. On the other side, the robot mirrors the motion of the stylus and its motion is displayed to the human operator via a video feed. Open and close commands to the gripper are sent using buttons on the stylus. Serial communication using Python serial library is used to control the gripper. Domain B is the network core which handles communication of packets between Domains A and C. It consists of Time-Sensitive Networking (TSN) capable switches (\textit{Edge Switch 1} and \textit{2}) that use P4 programmable Netronome Agilio SmartNICs. These SmartNICs are programmed to support Time Aware Shaper (TAS) which provides bounded latencies for time-critical Scheduled Traffic (ST) flows such as the haptic flow.


The task we have chosen to perform is to grip a tool using the gripper on the robot and to move it along the paths resembling specific shapes. In this work, we consider a rigid tool with a handle grip and consisting of a flat surface. We restrict the applied force to $75\%$ of the maximum. The grip is considered good if it is held firmly along the axis over its length. The two-fingered gripper's orientation is manipulated only in a plane parallel and equidistant to the gripping surfaces of the fingers.

\subsection{Building a Force Sensor Array}
When object grasping is enabled with a two-fingered gripper, force and contact region measurement can be achieved by placing a force sensor array on the gripper's fingertips. We built a $5 \times3$ force sensor array that covers the contact surfaces uniformly using force-sensitive resistors (FSR) which are inexpensive and widely available. The measurements from the FSR array are used to calculate pose correction as described in section \ref{PoseCorr}. The circuitry is made using a custom-built PCB with a footprint of $30~mm \times 40~mm$. The output from each FSR is connected to the 10-bit ADC input of an Arduino Mega2560Rev3 microcontroller. Arduino's internal voltage reference is enabled. The digital output is communicated to \textit{Host 2} over a serial port. The output value is around $600$ when the force applied by the gripper on the tool is $75\%$ of the maximum.

\begin{figure}[ht]
\vspace{-5pt}
\centerline{
\includegraphics[width=0.4\textwidth]{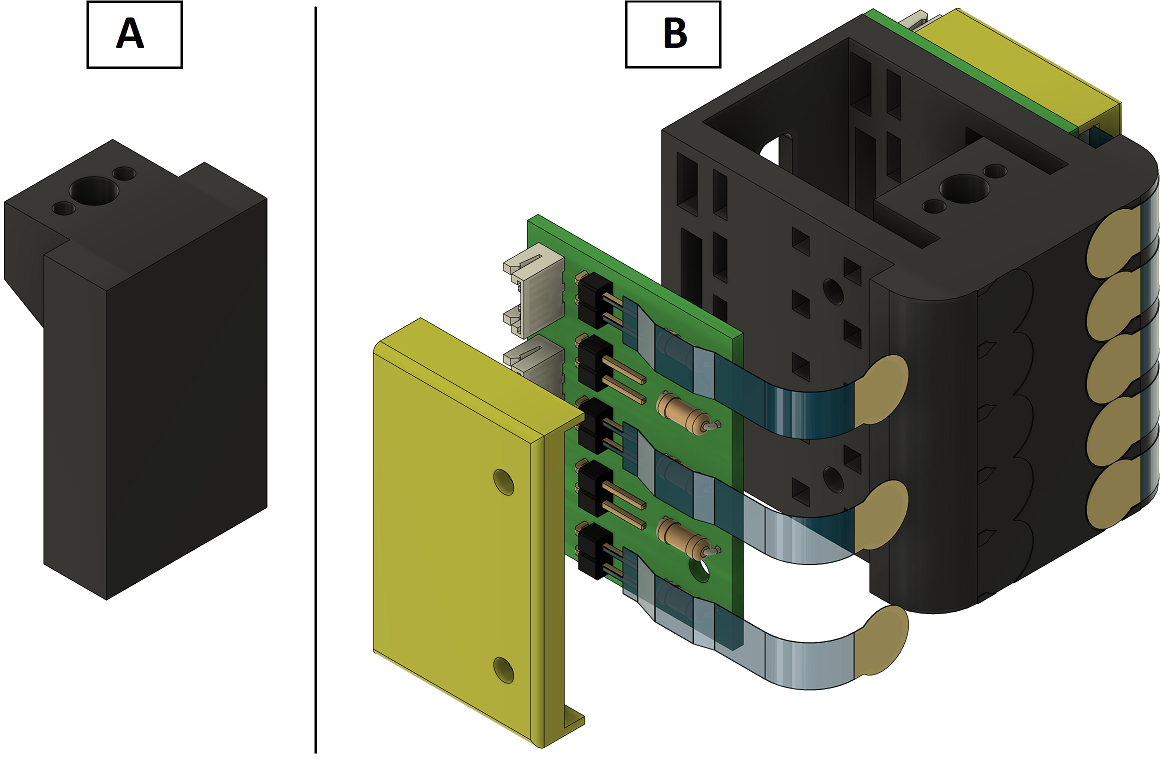}
}
\caption{CAD render of A.~stock and B.~modified gripper fingertips}
\label{fing-mod}
\end{figure}

The stock fingertip of the Robotiq 2F-85 gripper (Fig. \ref{fing-mod}A) is replaced with a custom fingertip (Fig. \ref{fing-mod}B). It is designed using CAD and fabricated using a 3D printer. This design allows for easy replacement of force-sensitive resistors (FSRs) and accommodates the PCB and the associated cables without impeding the gripper's functionality. The custom fingertips with the FSR array are mounted onto the gripper as depicted in Fig. \ref{fsr-photo}.

\subsection{Standardizing Data Exchange}\label{std-data}
In our work, it is crucial to establish efficient reading and manipulation of data in host systems using C and Python, as well as in switches using P4 across the three domains (Fig. \ref{testbed}). P4 targets are built to process packet headers in real time and are generally not designed to work with their payload due to high resource requirements. However, to enable edge intelligence on switch ports, there is a need to modify the coordinate data in the payload. In our approach, we achieve payload data manipulation in P4 by maintaining a specific structure in which the data is packaged across all the devices in the three domains. This is done by declaring a custom header called \textit{coordinate\_metadata} which corresponds to the structure of the payload in  P4.

\begingroup
\setlength{\tabcolsep}{4pt} 
    \begin{table}[ht] \small
    \vspace{5pt}
    \caption{\textit{coordinate\_metadata} header}
    \begin{center}
        \begin{tabular}{|c|c|c|c|c|c|c|c|c|c|c|c|c|} 
        \hline
        SID & TID & x & y & z & qx & qy & .. & b1 & b2 & f0  & .. & f14\\
        \hline 
        \end{tabular}
    \label{header-format}
    \end{center}
    \vspace{-10pt}
    \end{table}
\endgroup

Table~\ref{header-format} shows the \textit{coordinate\_metadata} header, made up of different 16-bit fields. The data is sent using user datagram protocol (UDP) over TSN-enabled ethernet. The relevant data from any of the domains are inserted into  various fields of \textit{coordinate\_metadata}. Each UDP packet with the data of the haptic device, sent by \textit{Host 1}, has a unique sequence ID (SID). The task ID (TID) field is used to track different tasks and is explained in section \ref{TID}. The coordinates of the input haptic device are sent using x,y and z fields with sign information. The button information is sent using b1 and b2 fields. Feedback data from \textit{Host 2} contains the pose of the robotic arm, which comprises position and quaternion orientation, which is contained in the fields x, y, z, qx, qy, qz, and qw. The force sensor data is sent using the fields f0 - f14. This feedback data is used by the switch ports for pose correction. A telemetry header is also sent with the packet giving the ingress and egress timestamps, the packet length, and the ingress and egress ports. These attributes are used in performance evaluation. The total size of each ethernet frame is $130~bytes$.

The byte-ordering scheme of the Intel processors in \textit{Host 1} and \textit{Host 2} is Little-Endian while that of the SmartNIC is Big-Endian. Therefore, we convert each 16-bit field in \textit{coordinate\_metadata} into Big Endian at each host before transmitting it to the switch. Since P4 does not have floating-point numbers as a built-in base type \cite{P4_user_guide}, to transmit position and quaternion data from the hosts to the switch, we multiply them by appropriate exponents of ten and convert the fields to signed integers. We divide the received data by the same multiplier at the hosts. The multipliers are chosen such that the position resolutions of the haptic device and the robot are not constrained by data-type conversion.

\section{EDGE INTELLIGENT SWITCH PORTS} \label{int-port}

\subsection{Automatic Pose Correction} \label{PoseCorr}

\begin{figure}[ht]
\centerline{
\includegraphics[width=3.3in]{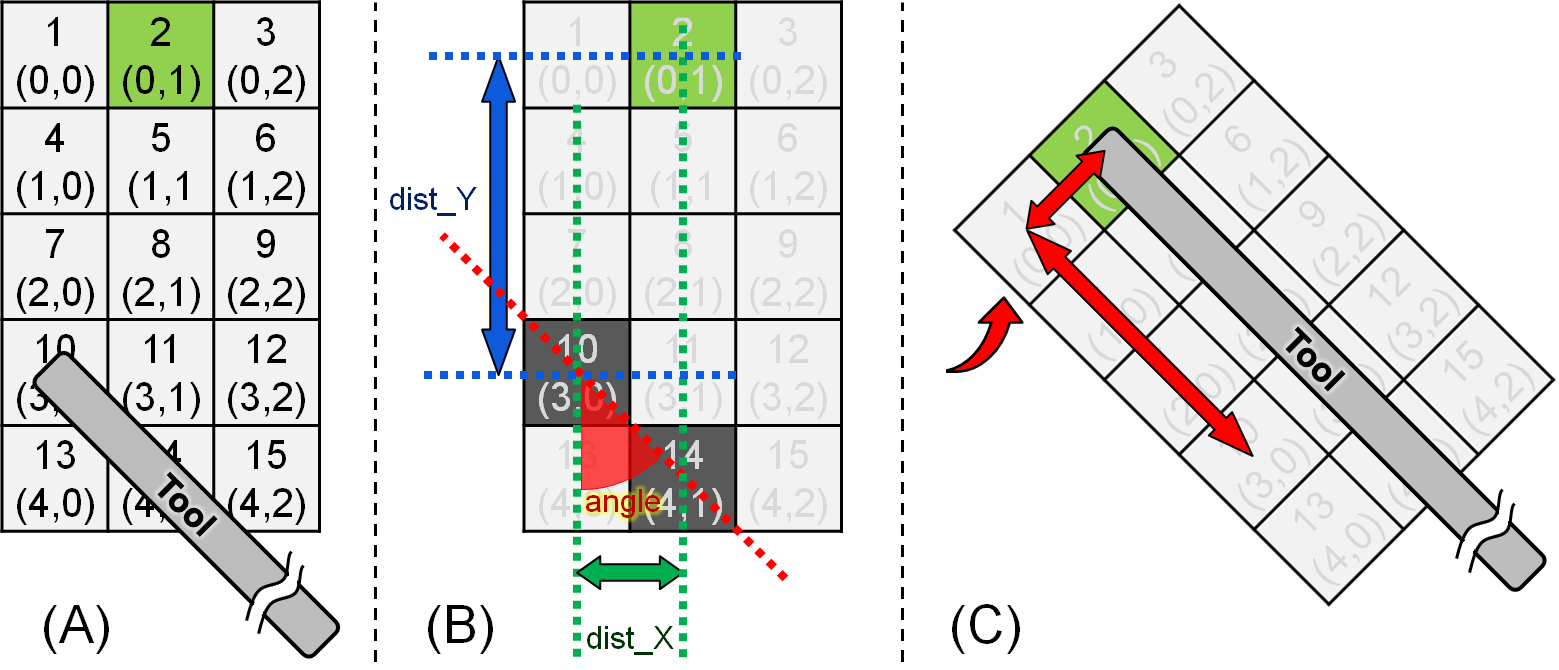}
}
\caption{An illustration of pose correction}
\label{45_deg}
\end{figure}

Algorithm~\ref{alg:pose-cor} outlines the pose correction method, which computes the robot's translation and rotation steps required for accurate gripping of the tool. This is a one-time calibration step for a specific tool and generates a baseline table of pose corrections for all possible orientations. The pose correction output is stored as a match action table in a \textit{json} file which is used by the P4 program running on the edge switch ports performing automatic pose correction. The corrected pose is obtained by calculating the translation distance and angle using basic cartesian coordinate arithmetic. This is demonstrated in Fig. \ref{45_deg}. In Fig.~\ref{45_deg}A, to begin with, the robot's gripper finger is positioned such that the tool is placed at the bottom at a certain angle. It presses against the force sensors 10 and 14, and for this position, Fig. \ref{45_deg}B shows the distance along the X and Y axes and the angle along which the robot has to move in order to achieve the correct orientation of the gripped tool. These distances and the angle are indicated by $dist_x$, $dist_y$ and \textit{angle}, respectively. The algorithm forms a list of all the combinations of sensor pairs and calculates $dist_x$, $dist_y$ and \textit{angle} with respect to Sensor 2~(shaded in green). The final gripper position after pose correction is shown in Fig. \ref{45_deg}C. When the tool is held in the correct position, the readings of the FSRs in the middle column will cross the threshold. If the gripper is closed without holding the tool, all the FSR readings exceed the threshold. These conditions are stored in the output.

    \begin{algorithm}[ht!] \small
    \caption{Pose correction calculation}
    \label{alg:pose-cor}
    \begin{algorithmic}[1]
         \renewcommand{\algorithmicrequire}{\textbf{Input:}}
         \renewcommand{\algorithmicensure}{\textbf{Output:}}
         \REQUIRE $n$, $m$, $gap_x$, $gap_y$ 
         \ENSURE  correction table         
         \STATE Create an array of $n\times m$ FSRs
         \comment{Sensor array has $n\times m$ FSRs spaced apart by $gap_x$ and $gap_y$ in horizontal and vertical axes (Fig. \ref{45_deg}A)}
         \STATE For every FSR pair, execute steps 3 \& 4
         \comment{Example calculations below for cells 10 and 14 (Fig. \ref{45_deg}B)}
         \STATE $dist_x = (x_{10} - x_2)\times gap_x$; $dist_y = (y_{10}-y_2)\times gap_y$;
         \comment{Calculate distance between base cell 2 (green) and chosen cell}
         \STATE $angle = (x_{10}-x_{14})\times gap_x / (y_{10}-y_{14})\times gap_x$
         \comment{Calculate the angle between the two FSRs and convert to quaternions $qx$, $qy$, $qz$, $qw$ using Euler angles to quaternions conversion formula}
         \STATE Store $dist_x$, $dist_y$, $quaternions$ ($qx$, $qy$, $qz$, $qw$) in a correction table
     \end{algorithmic} 
    \end{algorithm}

\begingroup
    \setlength{\tabcolsep}{4pt} 
    \begin{table}[ht!] \small
    \vspace{5pt}
    \caption{Example force sensor array reading}
    \begin{center}
    \begin{tabular}{|c|c|c|c|c|c|c|c|c|c|c|c|} 
     \hline
     FSR array & 3 & 5 & .. & 8 & 20 & \textbf{720} & 12 & 0 & 1 & \textbf{760} & 3\\ \hline
     \textit{Index} & \textit{0} & \textit{1} & .. & \textit{7} & \textit{8} & \textit{9} & \textit{10} & \textit{11} & \textit{12} & \textit{13} & \textit{14}\\
     \hline 
    \end{tabular}
    \label{SeEsExample}
    \end{center}
    \vspace{-10pt}
    \end{table}
\endgroup

\begin{figure}[ht]
\vspace{-5pt}
\centerline{
\includegraphics[width=3.3in]{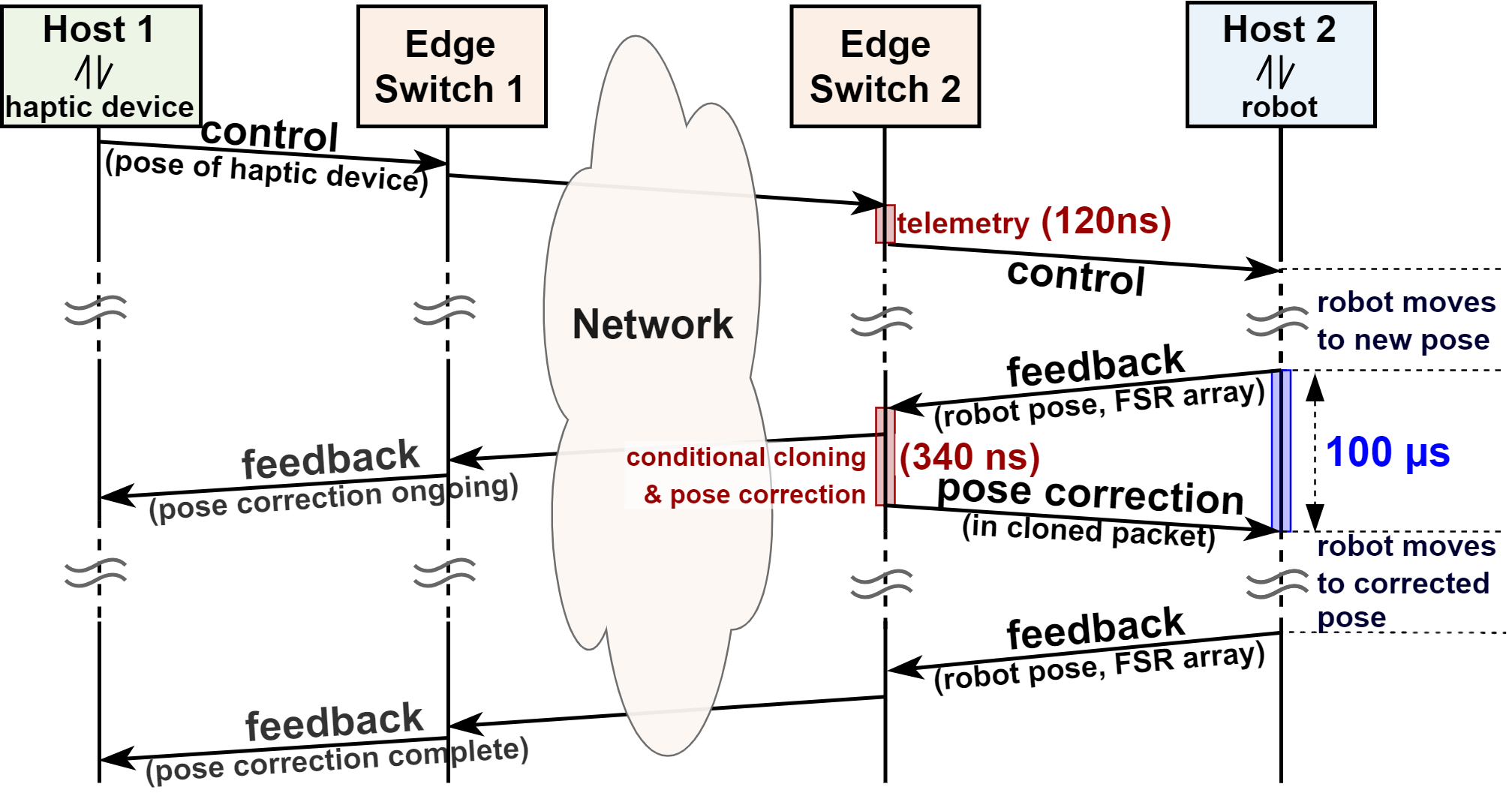}
}
\caption{Automatic pose correction - packet flow}
\label{ladder}
\end{figure}

Fig. \ref{ladder} shows the entities in the network including \textit{Host 1} interfaces the haptic device, \textit{Host 2} which is equipped with the robotic arm and the edge switch which is accessible to \textit{Host 2}. It illustrates the flow of control, feedback and pose correction packets. 
The haptic device originates control packets via \textit{Host 1} and the egress port of \textit{Edge Switch 2}  adds telemetry data to these before forwarding these to the teleoperator via \textit{Host 2}. Algorithm \ref{alg:PoseCorrP4} describes  implementation of grip inspection and pose correction intelligence in the switch port using P4. Feedback packets from \textit{Host 2} containing the pose of the robot and readings of the force sensor array in \textit{coordinate\_metadata} header are inspected by the ingress port of \textit{Edge Switch 2}. 
The P4 program begins by parsing the \textit{coordinate\_metadata} header. It then executes Algorithm \ref{alg:EdgeSensors}, an extern function developed in MicroC which identifies the sensors in the force sensor array exceeding a predefined threshold of $500$. The extern function also stores the indices of the first and the last among the identified sensors in registers named $se$ and $es$. Table \ref{SeEsExample} gives a snapshot of the \textit{coordinate\_metadata} header corresponding to force sensor array where $se$ is $9$ and $es$ is $13$. If $es$ is greater than $se$, indicating an incorrect grip, \textit{Edge Switch 2}  clones the packet before  forwarding it to the haptic device. The algorithm retrieves the necessary data ($dist_x$, $dist_y$, $quaternions$) from the match action table of precalculated pose corrections using the key $(se, es)$. These corrections are updated in the \textit{coordinate\_metadata} header of the cloned packet and the corrected packet is sent back to the teleoperator. The subsequent feedback packet from the teleoperator is inspected and the TID field in \textit{coordinate\_metadata} header is used to indicate whether the grip is correct or not, and the haptic device uses this information to notify the human operator. During regular operation of the teleoperator, the feedback packet is forwarded to the haptic device without cloning.

    \begin{algorithm}[ht] \small
    \caption{Grip inspection and pose correction implementation in P4}
    \label{alg:PoseCorrP4}
    \begin{algorithmic}[1]
    \renewcommand{\algorithmicrequire}{\textbf{Input:}}
    \renewcommand{\algorithmicensure}{\textbf{Output:}}
        \REQUIRE Teleoperator feedback (Fig \ref{ladder}) with \textit{coordinate\_metadata} header
        \ENSURE  a. Forward packet to \textit{Host 1} (feedback)\\b. Send packet with corrected pose in \textit{coordinate\_metadata} to \textit{Host 2}
        \STATE Parse coordinate\_metadata header
        \STATE Run extern function \textit{EdgeSensors} (Algorithm \ref{alg:EdgeSensors})
        \IF {($es \leq se$)}              
            \STATE forward packet to \textit{Host 1}
        \ENDIF
        \comment{No incorrect grip detected, so send feedback packet to haptic device}
        \IF {($es > se$)}      
            \STATE Clone the packet
            \STATE Forward original packet to \textit{Host 1}
            \STATE $(dist_x, dist_y, quaternions) = match\_action\_table(se,es)$
            \STATE $x = x +dist_x$;  $y = y + dist_y$; update quaternions
            \comment{Update the pose in \textit{coordinate\_metadata}}
            \STATE Forward correction packet to \textit{Host 2}
        \ENDIF
        \comment{Incorrect grip, so clone \& send corrections to robot; send feedback to haptic device}
    \end{algorithmic} 
    \end{algorithm}

    \begin{algorithm}[ht] \small
    \caption{\textit{EdgeSensors} extern function in MicroC}
    \label{alg:EdgeSensors}
    \begin{algorithmic}[1]
    \renewcommand{\algorithmicrequire}{\textbf{Input:}}
    \renewcommand{\algorithmicensure}{\textbf{Output:}}
        \REQUIRE $FSR$, $threshold$
        \ENSURE  End indices of sensors in the array that exceed threshold stored in registers $se$, $es$
        \comment{Example: index $9$ \& $13$ in Table \ref{SeEsExample} corresponds to cells $10$ \& $14$ in Fig. \ref{45_deg}B}
        \comment{$se$: start to end; $es$: end to start}
        \FOR {$i = 0$ to $14$}
            \STATE $se = i$
            \IF{$FSR[i] > threshold$}
                \STATE break
            \ELSE
                \STATE $i++$
            \ENDIF
        \ENDFOR  
        \STATE Similarly get $es$ by looping in reverse
        \STATE Store $se$ \& $es$ in register memory
    \end{algorithmic}
    \end{algorithm} 

\subsubsection{Experimental Results and Discussion}
    
Fig. \ref{ladder} captures the time taken by the packets across different events in the edge intelligent switch port running the automatic pose correction algorithm. We include the ingress and egress timestamps of packets into the telemetry header. The difference between the two is used to measure the time a packet spends inside the switch. Our experiments reveal that adding timestamps and port information requires a maximum of $120~ns$. Furthermore, conditionally cloning a feedback packet and updating it with pose correction requires a maximum of $340~ns$. The time duration at \textit{Host 2} between sending a feedback packet and receiving the corresponding pose correction packet using edge intelligent switch port is observed to be less than $100~\mu s$. In contrast, in conventional pose correction, the feedback packets are inspected and  corrections are calculated at the source (\textit{Host 1}) and the control packets are sent back over the network. Not only this adds to the network load and latency, but the  packet losses across the network could indefinitely stall the pose correction and feedback process.

\begin{figure*}[t] 
    \centering
    \includegraphics[width=0.8\textwidth]{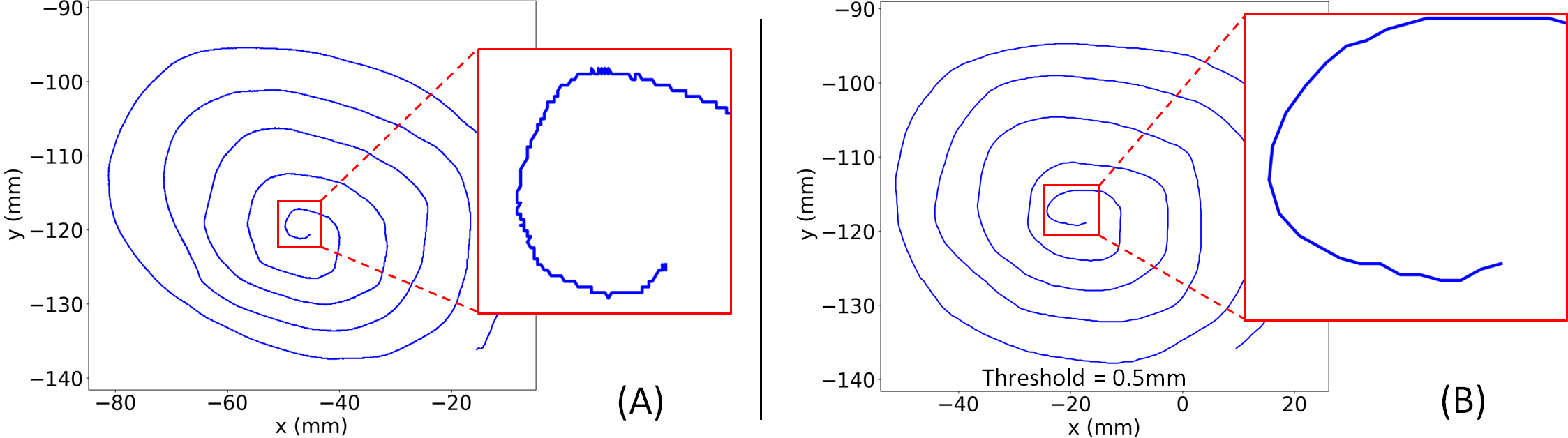}
    \caption{(A) Tremor reflected in output movement; (B) Smooth movement using threshold of 0.5mm}
    \label{trem_comparison} 
\end{figure*}

\subsection{Tremor suppression} \label{TremRed}
The hand tremors of the operator while manipulating the haptic device translate into vibration of the teleoperator. We use a tremor suppression algorithm implemented in P4 to prevent the haptic packets associated with tremors from being sent to the teleoperator. This is described in Algorithm~\ref{alg:trem-red} which is executed directly on the ingress port of  \textit{Edge Switch 1} (Fig.~\ref{testbed}). The algorithm begins with storing the initial 3D coordinates contained in the haptic data packet in the register memory. The coordinates in the subsequent haptic packet are compared to the stored values. If the $L_1$-distance between these coordinates is greater than a threshold, the packet is forwarded and the new coordinates are stored, else the latter haptic data packet is discarded.\footnote{More prevalent Euclidean distances are not used because P4 does not support {\it squaring}.} P4 does not have native support to compare negative numbers. Therefore, to manage the direction of tremors, we have developed a MicroC function supported directly on the switch port that converts negative values to absolute values.

    \begin{algorithm}[ht] \small
    \caption{Tremor suppression algorithm in P4-MicroC}
    \label{alg:trem-red}
    \begin{algorithmic}[1]
    \renewcommand{\algorithmicrequire}{\textbf{Input:}}
    \renewcommand{\algorithmicensure}{\textbf{Output:}}
        \REQUIRE  $thresh$, $x$, $y$, $z$ in \textit{coordinate\_metadata} in packets from \textit{Host 1}
        \ENSURE  Conditionally accept or discard haptic data packets
        \STATE $(diff_x \gets x - old_x)$; $(diff_y \gets y - old_y)$; $(diff_z \gets z - old_z)$;
        \IF {$(diff_x>>15 == 1)$} 
            \STATE ${diff_x \gets not(diff_x) +1; }$
        \ENDIF
        \comment{Handling negative numbers - get absolute value from 2's complement representation}
        \STATE Do the above for $diff_y$ and $diff_z$
        \IF {$diff_x + diff_y + diff_z > thresh$} 
            \STATE ${old_x \gets x; old_y \gets y; old_z \gets z;}$
        \comment{Update last stored coordinates}
            \STATE accept the packet
        \comment{Intentional motion $\implies$ forward data}
        \ELSE 
            \STATE drop the packet
        \ENDIF
    \end{algorithmic} 
    \end{algorithm}

\subsubsection{Experimental Results and Discussion}
We demonstrate our tremor suppression algorithm by executing standard tasks  such as a. Drawing a spiral using the stylus of the haptic device b. Drawing a straight line and c. Holding the stylus steadily~\cite{alty2017use}.  We record the coordinates in the received packets at \textit{Host 2} which are meant for the robot~(Fig.~\ref{testbed}). We experiment with the thresholds of $0.1~mm$, $0.5~mm$, $1~mm$ and $2~mm$. 
 The effect of the tremor suppression algorithm can be seen in Fig. \ref{trem_comparison}. Fig. \ref{trem_comparison}A shows the output movement of the teleoperator in absence of the suppression algorithm. The hand tremors captured by the stylus of the haptic device are clearly visible in the form of zig-zag motion of the teleoperator. Fig.~\ref{trem_comparison}B shows the smooth output resulting from application of our Algorithm~\ref{alg:trem-red}.

Tactile applications typically require control data to be sent at $1$~{\it KHz}. Our haptic device provides a $1$~\textit{KHz} signal with a minimum amplitude of $55~
\mu m$. Since the packet size is $130~bytes$ including the \textit{coordinate\_metadata} header, this results in a data transmission  rate of about $1~Mbps$. The tremor suppression algorithm with a threshold of $0.5~mm$ for the three tasks results in a reduced data rate as low as $55 Kbps$ with a bandwidth savings of over $94\%$. Table~\ref{Trem-Red-Result} captures the number of packets discarded and the overall savings in bandwidth. Since drawing a spiral is more involved compared to drawing a  straight line or holding the stylus, it warrants a higher number of packets to be transmitted. Consequently, the hold task has the maximum savings in bandwidth among the three. These tasks were performed in $15$, $5$ and $10$ seconds, respectively. The effect of setting different thresholds while drawing a spiral is shown in Table~\ref{Thresh-vary}. For example, for $2~mm$ threshold, the data transmission rate can be as low as $15~Kbps$. Furthermore, the last column shows that by suppressing tremors, we also reduce the total length travelled by the tool of the teleoperator. In a robot-assisted medical task such as making an incision, this can lead to reduced bruising.


\captionsetup[table]{skip=0pt} 
    \begin{table}[!ht] \small
    \vspace{5pt}
        \caption{Traffic reduction for different tasks}
        \centering
        \begin{tabular}{|P{0.05\textwidth}|P{0.085\textwidth}|P{0.075\textwidth}|P{0.065\textwidth}|P{0.06\textwidth}|}
        \hline
            \textbf{Task} & \textbf{No. of packets transmitted} & \textbf{No. of packets discarded} & \textbf{Average data rate (kbps)} & \textbf{Traffic reduction (\%) } \\ \hline
            Spiral & 813 & 14182 & 55 & 94.58  \\ \hline
            Line & 250 & 4718 & 51 & 94.97  \\ \hline
            Hold & 5 & 9950 & 0.5 & 99.95  \\ \hline
        \end{tabular}
    \label{Trem-Red-Result}
    \end{table}

    \begin{table}[!ht] \small
    \vspace{0pt}
        \caption{Effect of different thresholds in drawing a spiral}
        \centering
        \begin{tabular}{|P{0.05\textwidth}|P{0.075\textwidth}|P{0.065\textwidth}|P{0.055\textwidth}|P{0.05\textwidth}|P{0.04\textwidth}|}
        \hline
            \textbf{Thresh-old (mm)} & \textbf{No. of packets transmitted} & \textbf{No. of packets discarded} & \textbf{Average data rate (kbps)} & \textbf{Traffic reduction (\%)} & \textbf{Draw-ing length (mm) } \\ \hline
            None & 14992 & 0 & 1015 & 0.0 & 600  \\ \hline
            0.1 & 2299 & 12621 & 148 & 84.6 & 507  \\ \hline
            0.5 & 813 & 14182 & 55 & 94.6 & 527  \\ \hline
            1 & 423 & 14494 & 29 & 97.2 & 506  \\ \hline
            2 & 226 & 14714 & 15 & 98.5 & 512  \\ \hline
        \end{tabular}
        \label{Thresh-vary}
    \end{table}

\subsection{Seamless Switching Between Algorithms} \label{TID}
We use the 16 bit TID in the \textit{coordinate\_metadata} header not only to identify the capability of the robot but also to switch between different edge intelligence algorithms associated with different tasks. This framework facilitates edge intelligence algorithms to be offloaded to an edge switch that might connect to multiple robots. The necessary parameters linked to various tasks and robots are stored in the match action table of the switch. For example, $dist_x$, $dist_y$ and $quaternions$ are stored for a pose correction task, whereas the $threshold$ is stored for a tremor suppression task. When a robot connected to a particular switch port is required to perform a different task or is replaced with another robot, the new task's TID is used as a key, and the switching process is transparent to the end user. The robot is automatically assigned tasks without reprogramming the switch. The parameters can be changed in runtime by updating the match action table. Thus, P4-based edge intelligent switches can identify and track all connected robots and associated tasks, making the proposed solution highly flexible and scalable.

\section{CONCLUSION AND FUTURE WORK} \label{conclusion}
We have presented our hardware implementation of two edge intelligence algorithms, {\it pose correction} and {\it tremor suppression}, on P4 programmable switch ports using SmartNICs. The experimental results in Section~\ref{int-port} vindicate that introducing edge intelligence  can substantially reduce control loop latency and network load. Further, multiple algorithms can be hosted on the same  edge switch which can transparently switch between the algorithms depending on the tasks.

Physiological tremors are influenced by stress, anxiety, fatigue, caffeine intake, or medication side effects. So, there is scope to implement a dynamic tremor elimination algorithm.
Our future work entails designing and implementing more advanced edge intelligence algorithms, especially ones using artificial intelligence, covering a wide class of tasks. 
We can also explore security as a feature in the edge intelligent switches.

\section*{Acknowledgments}

This work was partly supported by the Ministry of Electronics and Information Technology, Government of India (SP/MITO-20-0006) and partly by the Centre for Networked Intelligence (a Cisco CSR initiative) at Indian Institute of Science, Bengaluru, India.

\bibliographystyle{unsrt} 
\begin{small}
\bibliography{references
}
\end{small}


\end{document}